\pdfoutput=1
\pdfoutput=1
\pdfoutput=1
\pdfoutput=1
\pdfoutput=1
\documentclass[9pt,letterpaper,twocolumn]{article} 

\usepackage{ol}
\usepackage{hyperref}
\usepackage{amsmath}

\begin{document}

\twocolumn[ 

\title{Implementation of conformal digital metasurfaces for THz polarimetric sensing}

\author{Javad Shabanpour, Sina Beyraghi, Fardin Ghorbani, and Homayoon Oraizi}

\address{Iran University of Science and Technology, Tehran, Iran

Email:m.javadshabanpour1372@gmail.com}


\begin{abstract}
Monitoring and controlling the state of polarization of electromagnetic waves is of significant interest for various basic and practical applications such as linear position sensing and medical imaging. Here, we propose the first conformal digital metamaterial absorber to detect the polarization state of THz incident waves. The proposed polarimeter is capable of characterizing four independent polarization states of (TE, TM, $\pm {45^\circ}$ linear, and RCP/LCP) by observing the reflectivity of the structure with respect to the x- and y-direction. Besides, the proposed structure displays a strong absorptivity above 90\% up to the incidence angle of $50^{\circ}$ for oblique incident waves with different polarizations. By mere changing the bias voltage of two orthogonal VO2 microwires via two independent computer-programmed multichannel DC network, distinct conditions for reflected waves occurs under excitations of different polarizations, whereby the polarization state of the incident wave may readily be estimated. We believe that the proposed metasurface-based polarimeter can pave the way for polarization detection applications on curved surfaces.  
\end{abstract} 


 ] 

\noindent
 In recent years,  Terahertz (THz) waves have reached maturity and have witnessed remarkable attention due to their distinct features that make them beneficial in various sensing and imaging technologies\cite{1}. Artificial metasurfaces, as periodic arrangements of engineered sub-wavelength structures, can be structured for complex and functional manipulation of electromagnetic (EM) waves\cite{9}. They can be exploited to modify the permittivity and permeability of materials by purposefully controlling the local properties of phase, amplitude, and polarization of EM waves. Beyond the scope of analog metamaterials, digital metamaterials have experienced rapid evolution compared to traditional wave manipulations and outstandingly provide a broader range of wave-matter functionalities which make them fascinating in various practical applications such as smart surfaces, power intensity control, and programmable structures\cite{2,3,17}.
\begin{figure*}
	\centering
	\includegraphics[height=8cm]{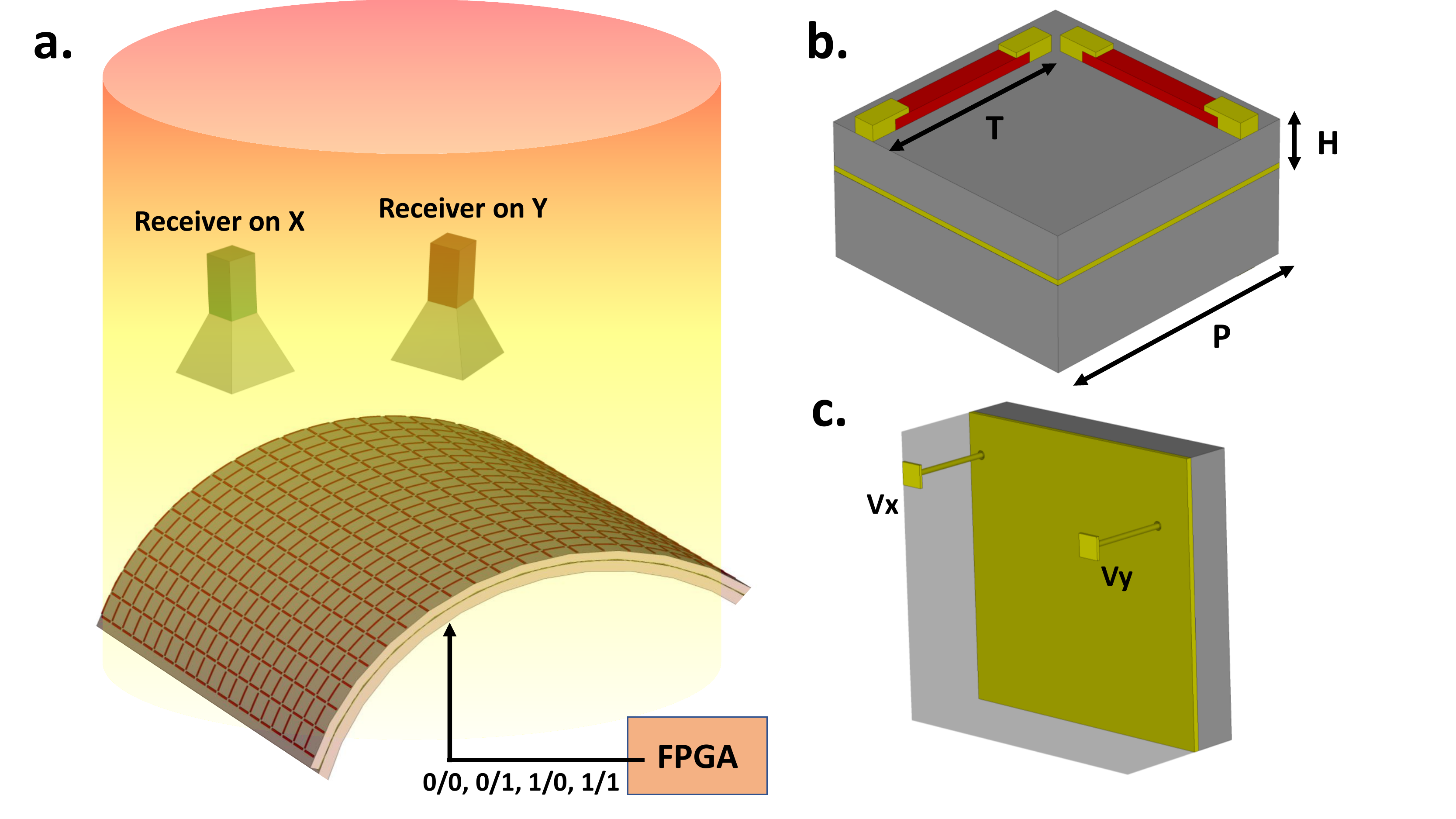}
	\caption{ a) The schematic diagram of conformal metamaterial absorber for polarization detection application. (b), schematic view of the unit-cell. (c) schematic view of the back of the unit-cell where Vx and Vy are two patches that serve as positive electrodes for applying DC bias voltage. }
	\label{fgr:example2col}			
\end{figure*}

Given that each polarization can represent an independent information in communication systems, the polarization can be considered as one of the fundamental attributes characterizing  the features of EM waves, and the measurement of polarization divulges rich information about the nature of scattering, radiation, and absorption phenomena. A polarization-dependent metamaterial absorber is an ideal building block for polarimetric sensing systems which can equip an inspiring platform for solving some crucial THz challenges such as linear position sensing, security scanning, and medical imaging\cite{4,5,16}.

After introducing the first perfect metamaterial absorber (MMA) by Landy et al. in 2008\cite{6}, as a counterpart to conventional absorbers, MMAs have become research hotspots and have also been of great attention for solving EM interference challenges, such as the stealth technologies, radar absorbing and thermal emitter\cite{7}. Generally, the maximum absorption is obtained for the normal incidence. However, when a metamaterial absorber is built on a curved surface, its absorption efficiency drops drastically. Accordingly, the flexibility and incident angle insensitivity are two main factors for designing MMAs for curved surfaces that have been followed by some works in recent years\cite{8}.

The necessity of polarimetric sensors for curved surfaces has led us to design a first conformal MMA for detection of the polarization states of THz wavefronts. Our proposed phase/polarization encoded anisotropic digital metasurface can exhibit the perfect absorptivity in wide incident angles based on tuning the proper electrical resistivity of vanadium dioxide (VO2).

VO2 is a well-known tunable material that undergoes an ultrafast reversible phase transition from insulator monoclinic to metallic tetragonal phase above the critical temperature around ${{\rm{T}}_{\rm{c}}}{\rm{ = 340K}}$. This metal-insulator transition (MIT) can be provoked by thermal, optical, or electrical activation which can occur within an order of several nanoseconds or even in picoseconds range for optical stimuli\cite{10}. Due to the dramatic changes of electrical and optical properties of VO2 across the two phases, it has been identified as a notable substance in tunable metamaterial devices over a broad spectral range and has myriad applications in the THz regions, such as THz waves modulator, angle-insensitive absorber, and programmable ultrafast meta-devices\cite{11}.

In this paper, a THz polarimetry device is proposed as depicted in \textbf{Fig. 1}, which comprises two perpendicular VO2 microwires whose reflection characteristics can be dynamically tuned in a real-time manner independently for two orthogonal x- and y-polarized excitations. The proposed VO2- integrated anisotropic metasurface (VIAM) can retain the perfect absorptivity in wide oblique incidences for THz waves that make them suitable for a curved surface. By digitally changing the spatial voltage distribution provided by two independent multichannel DC bias voltages, and observing the reflection properties of scattered beams via two receiving antennas, the state of incident polarization can be determined. To demonstrate
the main mechanism of VIAM, the relationship between spatial coding sequences and the polarization of the incident beam is investigated. Besides, by analyzing the induced electric field and the power loss density of the proposed VIAM, we show that increasing the electrical conductivity of VO2 plays an effective role in dissipating the incoming EM energy especially when the oblique incident angle increases. To the best knowledge of the authors, our proposed VIAM is the first conformal polarimetric device. We believe that the integrated structural design, ultrafast switching time, and wide coverage of incident-angle characteristics pave the way for many scientific and engineering applications, such as linear position sensing and future practical THz devices.

\textbf{Fig. 1b} shows the sketch representation of the proposed meta-particle composed of four layers of VO2 microwires, polyamide, gold plate, and polyimide from top to bottom respectively. Two pieces of separated patches are located beneath the bottom substrate as positive electrodes which are independently connected to two DC bias voltage networks as shown in \textbf{Fig. 1c}. The gold plate acts as a negative electrode which is connected to two lateral metalized electrodes for each VO2 microwires through two metallic via holes. Through this mechanism, all the biasing wires and networks can be located beneath the structure without interfering with the EM properties of the metasurface. The geometrical dimensions are P=50$\mu m$,H=17$\mu m$ and T=40$\mu m$ respectively. The thickness and the width of the VO2 microwires are 1$\mu m$ and 4$\mu m$.

Typical VO2 thin films show electrical conductivity in the range of $10 \sim 100\,\,{\rm{S/m}}$ with the relative permittivity of about 9 and after occurring the structural transformation, its electrical conductivity reaches as high as an order of ${10^5}$ in the metallic state\cite{12}. We have performed full-wave simulations by the commercial CST software. For the evaluation of reflection properties of the infinite array of VIAM, the periodic boundary conditions are utilized along the x and y directions and the open boundary condition is employed along the z-axis. Meanwhile, linearly and circularly polarized plane waves at different incidence angles illuminate the VIAM structure.
\begin{figure*}
	\centering
	\includegraphics[height=9cm]{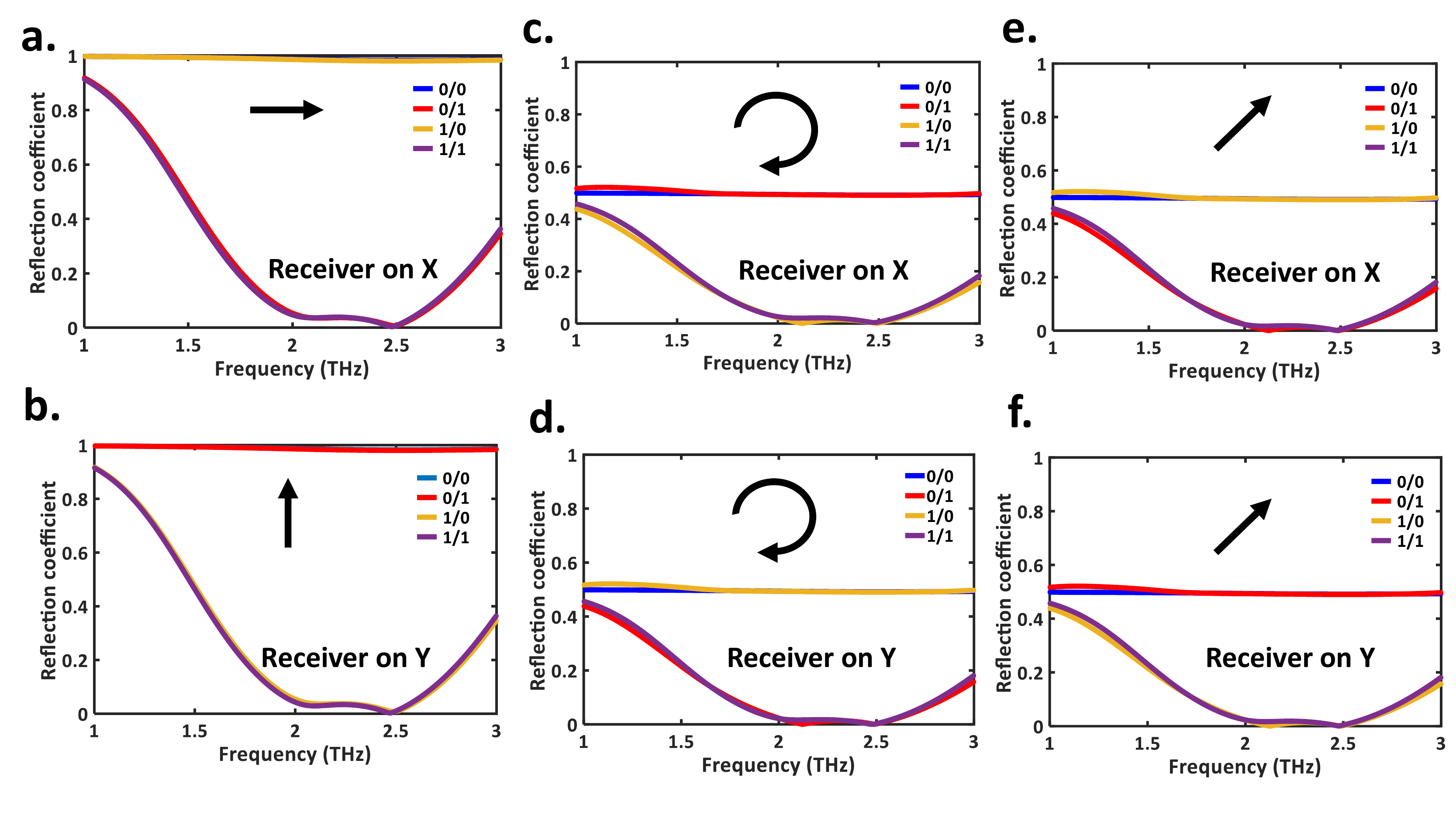}
	\caption{ Simulated results of reflectivity of the designed VIAM structure under the excitation of a) TM polarized THz incident waves. b) TE polarized THz incident waves. c,d) Circular polarized THz incident waves for receiving antenna on X and Y directions respectively. e,f) $+45^\circ$ linear polarized THz incident waves for receiving antenna on X and Y directions respectively.}
	\label{fgr:example2col}			
\end{figure*}

Since the thickness of a ground plane (gold layer) is much larger than the penetration depth of the incident THz wavefronts ($T(\omega ) = 0$), the absorptivity of the VIAM can be calculated by Eq. 1, where $A(\omega )$ and $\Gamma (\omega )$ are the absorption and reflectance, respectively\cite{13}.
\begin{equation}
A(\omega ) = 1 - \Gamma (\omega )
\end{equation}
When VO2 is in the dielectric steady-state phase and having the lowest electrical conductivity, the reflection characteristics of the VIAM is in its highest value. On the other hand, at the intermediate temperatures, the transmitted waves are dissipated due to the significant VO2 ohmic losses, and the maximum absorption efficiency is achieved. Therefore, we set ${\sigma _{off}} = 10S/m$ ( "0" digital state) and ${\sigma _{on}} = 5 \times {10^4}S/m$ ( "1" digital state). 
The conductivity of VO2 microwires along the x and y-direction are marked by ${\sigma _x}$ and ${\sigma _y}$, and by independently changing the DC bias voltage, four digital states of 0/0, 0/1, 1/0, and 1/1 correspond to ${\sigma _{off}}/{\sigma _{off}}$, ${\sigma _{off}}/{\sigma _{on}}$, ${\sigma _{on}}/{\sigma _{off}}$, ${\sigma _{on}}/{\sigma _{on}}$  will be attained where the binary codes before and
after the slash symbol ( / ) indicates the digital codes for the VO2 microwires along the x and y-directions, respectively.
\renewcommand{\figurename}{Table}
\renewcommand{\thefigure}{1} 
\begin{figure}[htb]
	\caption{Reflectivity for four digital states at the frequency of 2.4 THz and the corresponding distinct polarization states.}
	\centerline{\includegraphics[width=7cm]{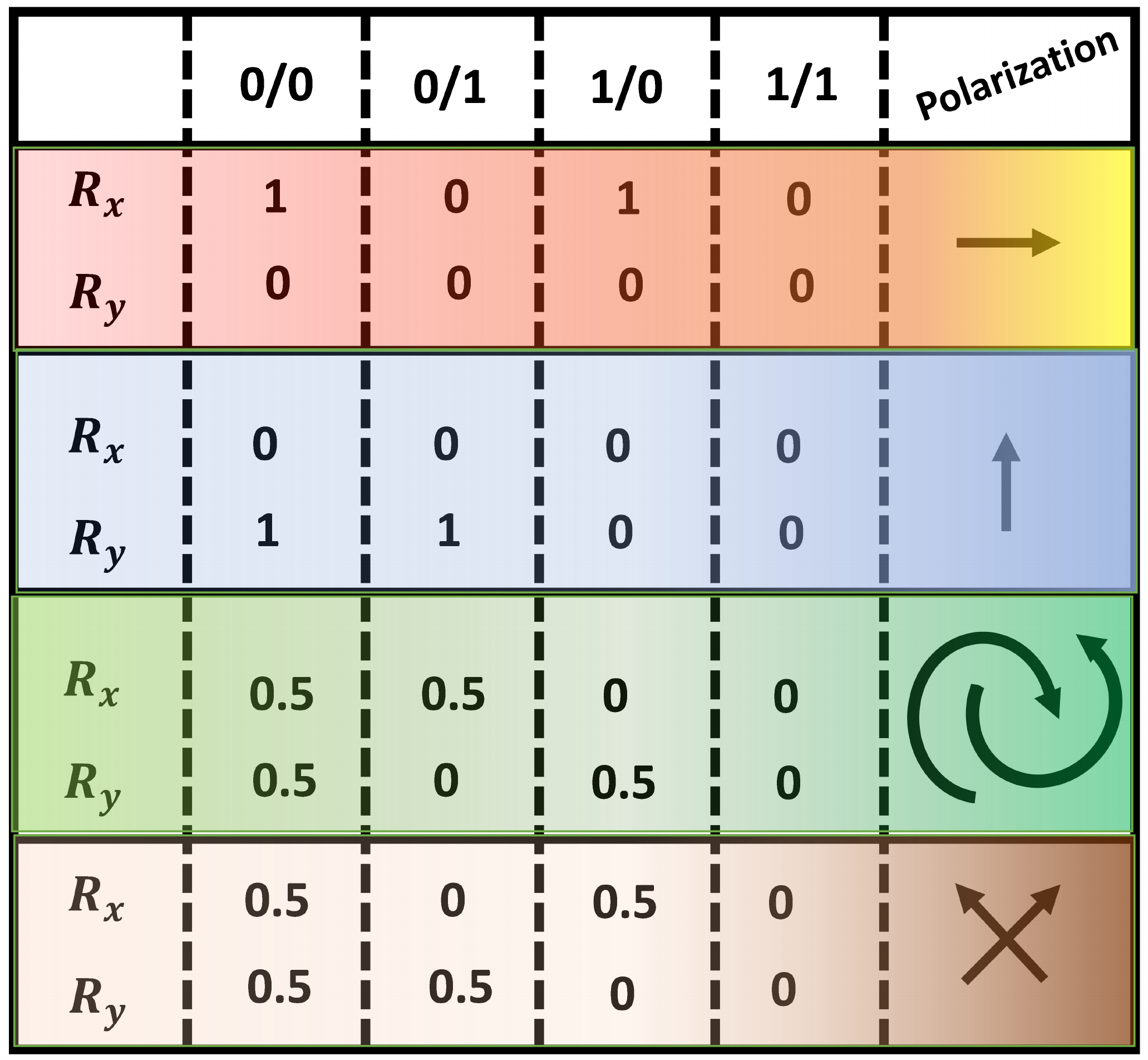}}
\end{figure}

Under the normal incidence, numerical simulations have been conducted for different polarized incident waves, and the reflection spectra of the proposed VIAM for four different digital states are illustrated in \textbf{Fig. 2}. The absorptivity of VIAM is novelty employed to characterize the four independent polarization states (TE, TM, $\pm {45^\circ}$ linear, and RCP/LCP). Consequently, the proposed structure demonstrates a wideband polarization-dependence of absorption for distinct independent polarization, thus, the high contrast can be realized for sensing applications. Consider the TM polarized waves as an example (depicted in the first row of \textbf{Table. 1}), since the VIAM has anisotropic structural nature, just those VO2 films which are parallel to x-direction are excited and the reflection amplitude response for cross-polarization is zero. Therefore, the receiving antenna towards the y-direction receives nothing for all the four spatial code distributions. Observe in \textbf{Table. 1} that the polarization cross-talk between two orthogonal directions is negligible, and changing the digital state of VO2 microwire in one orthogonal direction does not affect the electromagnetic response in the other direction for the LP incident THz wave. 

For another example, consider RCP THz incident waves illuminating the structure. For a digital state of 0/0, all the EM energy reflected by RCP, and according to the relationship between CP and LP polarization, we observe that half of the EM radiation reflected towards x-direction, and the remaining EM energy reflected towards y-direction. For the digital state of 1/1, all the incident THz wave energy is effectively absorbed by the proposed VIAM due to the ohmic losses of VO2 films. For the digital states of 0/1, a quarter of the incident wavefronts is reflected by RCP, and the other quarter of the wave energy is reflected by LCP, and these two wavefronts have the same phase difference. Therefore, half of the incident THz wavefront is reflected towards x-direction and the remaining wave energy is absorbed by the structure. In the same way, for the 1/0 digital state, half of the EM energy absorb by the structure and the remaining THz wavefront is reflected toward y-direction. As a final comment, all the conditions presented in \textbf{Table. 1} are distinct, and by observing the reflected energies through x- and y-polarization, one can readily determine the state of the incident THz wavefronts polarization. It should be noted that due to the nature of the structure, our designed VIAM can not differentiate between LCP and RCP as well as $+45^\circ$ LP and $-45^\circ$ LP. 
\renewcommand{\figurename}{Fig}
\renewcommand{\thefigure}{3} 
\begin{figure}[htb]
	\centerline{\includegraphics[width=7cm]{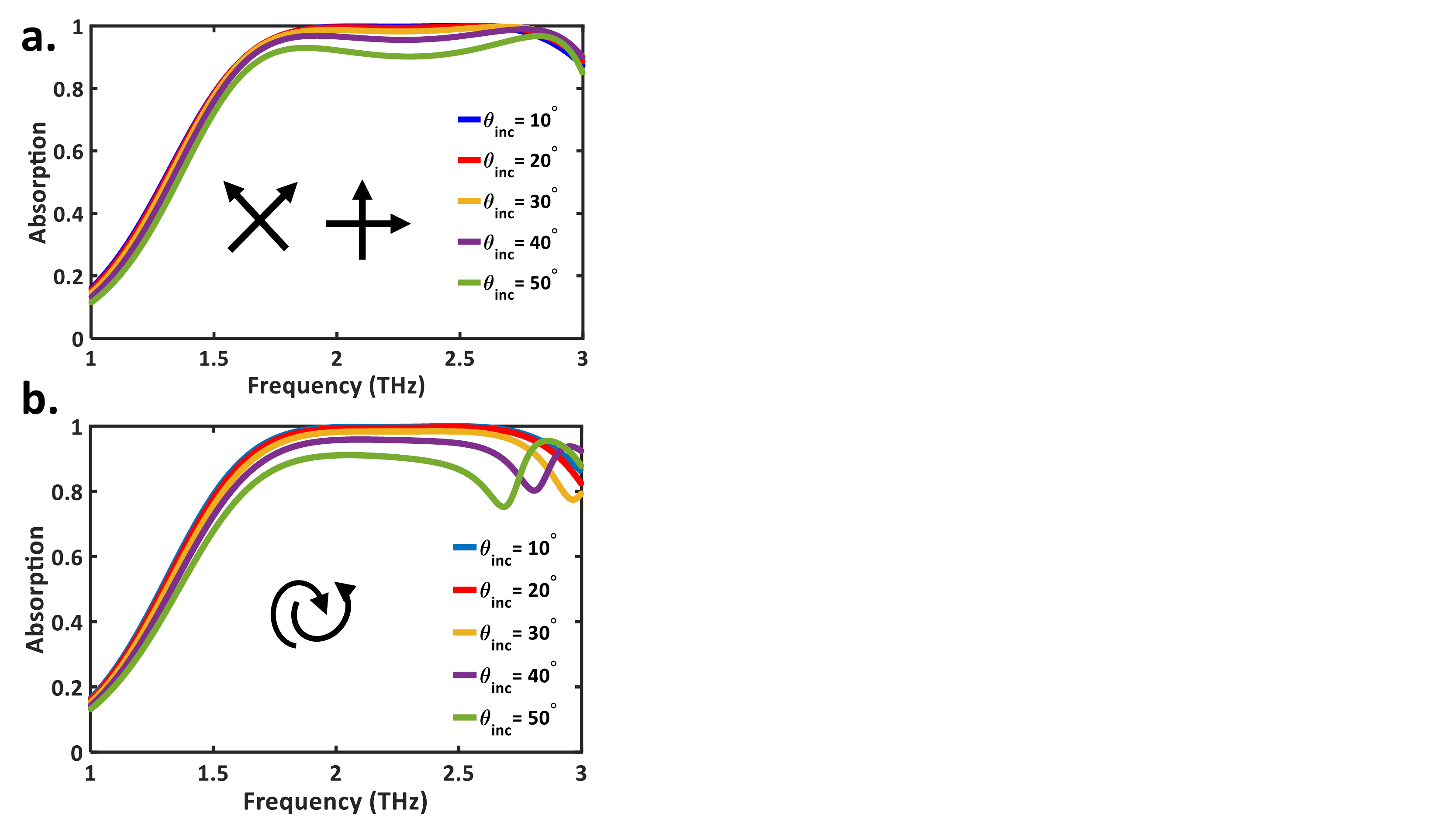}}
	\caption{The simulated result of absorptivity for different cases of a) (TE, TM, $\pm {45^\circ}$ linear)-polarized oblique incident THz wavefronts. b) RCP and LCP oblique incident THz wavefronts, when the electrical conductivity of VO2 microwires are 50000 S/m.  }
\end{figure}

When an MMA is fabricated on a curved surface, only those meta-atoms at the center of the absorber are illuminated at normal incidence, while, the other meta-particles on the curved surface are illuminated at oblique incidence. However, by increasing the incident angle, the degradation of absorptivity is inevitable\cite{14}. For instance, under TE and TM oblique incidence, the reflection coefficients can be obtained by\cite{15}:
\begin{equation}
{\Gamma _{{\rm{TE}}}}(\omega ) = \frac{{Z(\omega )\cos {\theta _i} - {Z_0}\cos {\theta _t}}}{{Z(\omega )\cos {\theta _i} + {Z_0}\cos {\theta _t}}}
\end{equation}
\begin{equation}
{\Gamma _{{\rm{TM}}}}(\omega ) = \frac{{Z(\omega )\cos {\theta _t} - {Z_0}\cos {\theta _i}}}{{Z(\omega )\cos {\theta _t} + {Z_0}\cos {\theta _i}}}
\end{equation}
where ${Z(\omega )}$ and ${{Z_0}}$ indicate the impedances of the VIAM and free space, and ${{\theta _i}}$ and ${{\theta _t}}$ are the incident and transmission angles, respectively. We will show that the near-unity absorption up to $50^{\circ}$ incident angles which is crucial for polarization detection applications can be achieved with our elaborately designed VIAM structure. The absorption spectra of the VIAM under different incident angles up to 50 degrees for (TE, TM, $\pm {45^\circ}$ linear)-polarized THz wavefronts is depicted in \textbf{Fig. 3a}. All the numerical simulations are conducted for the electrical conductivity of VO2 microwires equal to ${\sigma _{on}} = 5 \times {10^4}$. 
To better understand the main mechanism of absorption, notice that by increasing the incident angle, according to Eq. 4, the power loss density decreases for the electrical conductivity of VO2 in the off-state. 
\begin{equation}
{P_{loss}} = \int {\sigma {{\left| E \right|}^2}dv} 
\end{equation}
To compensate for this decline, the only way is to increase the electrical conductivity of VO2 microwires to reach the value of on-state by changing the DC bias voltage. Therefore, the power loss is mostly due to the ohmic losses of VO2 microwires, and changing the electrical conductivity of VO2 plays a beneficial role in dissipating the incoming EM energy.
By increasing the incident angle of THz incident wavefronts, the absorbing bandwidth becomes narrow, and our designed VIAM can retain its absorptivity up to $50^{\circ}$ in the band of 1.6-3 THz. Accordingly, our proposed structure shows a wideband absorptivity above 90\% in 1.6–3 THz for oblique incidences up to $50^{\circ}$ for different cases of incident polarization (LP, $\pm {45^\circ}$ linear, and CP (See \textbf{Fig. 3b})). 

In summary, we have introduced the first conformal metamaterial absorber for polarimetric detection that can characterize the four independent polarization states of (TE, TM, $\pm {45^\circ}$ linear, and RCP/LCP). Our proposed VIAM structure displays a strong absorptivity above 90\% up to the incidence angle of $50^{\circ}$. By encoding the VIAM in four distinct digital states through changing the independent DC bias voltages for two orthogonal VO2 microwires, our designed structure is successfully characterized the polarization of THz incident wavefronts with an ultrafast switching time between distinct digital states. We believe that the presented conformal metasurface-based polarimeters may find great potentials for solving some crucial THz challenges, such as linear position sensing, security scanning, and medical imaging.

\bigskip


\end{document}